\documentclass[aps,floats,floatfix,showpacs,preprintnumbers,amssymb,prd,twocolumn,superscriptaddress,nofootinbib,nolongbibliography,reprint]{revtex4-1}

\usepackage{amssymb,amsmath,verbatim,mathtools,needspace,enumitem,etoolbox,graphicx,physics,microtype,afterpage,xspace,tabularx,lmodern,multirow,boldline, makecell, booktabs}

\usepackage{appendix}
\usepackage{tensor}
\usepackage{array}
\usepackage{siunitx}
\usepackage[normalem]{ulem}
\usepackage[dvipsnames, usenames]{xcolor}
\usepackage{xr-hyper}
\definecolor{linkcolor}{rgb}{0.0,0.3,0.5}
\usepackage[unicode, colorlinks=true, linkcolor=linkcolor, citecolor=linkcolor, filecolor=linkcolor, urlcolor=linkcolor, linktocpage, breaklinks]{hyperref}
\usepackage[all]{hypcap}
\usepackage[T1]{fontenc}
\usepackage[utf8]{inputenc}
\usepackage[usenames,dvipsnames]{xcolor}
\definecolor{cornellGreen}{HTML}{6EB43F}
\definecolor{cornellRed}{HTML}{B31B1B}
\hypersetup{colorlinks=true,citecolor=cornellRed,linkcolor=cornellRed,urlcolor=cornellRed}
\usepackage{multirow}

\definecolor{romared}{RGB}{142,0,28}

\newcommand{\be}{\begin{equation}}
\newcommand{\ee}{\end{equation}}

\def\be{\begin{equation}}
\def\ee{\end{equation}}
\newcommand{\beq}{\begin{eqnarray}}
\newcommand{\eeq}{\end{eqnarray}}

\usepackage{makecell}
\usepackage{soul}

\newcolumntype{Y}{>{\centering\arraybackslash}X}

\newcommand{\barh}{\bar{h}}
 \definecolor{mypurple}{RGB}{130, 0, 130} 

\begin{document}
\title{Environmental effects in extreme mass ratio inspirals: \\  perturbations to the environment in Kerr}
\author{Conor Dyson}
\affiliation{Niels Bohr International Academy, Niels Bohr Institute, Blegdamsvej 17, 2100 Copenhagen, Denmark}
\author{Thomas F.M.~Spieksma}
\affiliation{Niels Bohr International Academy, Niels Bohr Institute, Blegdamsvej 17, 2100 Copenhagen, Denmark}
\author{Richard Brito}
\affiliation{CENTRA, Departamento de F\'{\i}sica, Instituto Superior T\'ecnico -- IST, Universidade de Lisboa -- UL, Avenida Rovisco Pais 1, 1049-001 Lisboa, Portugal}
\author{Maarten van de Meent}
\affiliation{Niels Bohr International Academy, Niels Bohr Institute, Blegdamsvej 17, 2100 Copenhagen, Denmark}
\affiliation{Max Planck Institute for Gravitational Physics (Albert Einstein Institute) 
Am Mühlenberg 1, D-14476 Potsdam, Germany}
\author{Sam Dolan}
\affiliation{Consortium for Fundamental Physics, School of Mathematical and Physical Sciences, University of Sheffield, Hicks Building, Hounsfield Road, Sheffield S3 7RH, United Kingdom.}
\date{\today}
\begin{abstract}
Future gravitational wave observatories open a unique avenue to study the environments surrounding black holes. Intermediate or extreme mass ratio inspirals will spend thousands to millions of cycles in the sensitivity range of detectors, allowing subtle environmental effects to accumulate in the gravitational waveform. Working in Lorenz gauge and considering equatorial circular orbits, we present the first self-consistent, fully relativistic calculation of a perturbation to a black hole environment due to an inspiraling secondary in the Kerr geometry. As an example case, we consider the environment to be that of a superradiantly grown scalar cloud, though our framework is generalizable to other scenarios. We demonstrate that the scalar field develops a rich wake structure induced by the secondary and compute scalar fluxes emitted to infinity and through the horizon. Relative differences in the fluxes compared to Schwarzschild are tens of percent on large intervals of parameter space, underscoring the importance of modeling in Kerr.
\end{abstract}
\maketitle
\noindent{\bf{\em Introduction.}}
Gravitational wave (GW) astronomy has progressed rapidly since the first detection of a binary black hole (BH) merger in 2015~\cite{LIGOScientific:2016aoc}, with over 100 events published to date~\cite{LIGOScientific:2018mvr,LIGOScientific:2020ibl,KAGRA:2021vkt}. Upcoming next-generation ground-based detectors, such as the Einstein Telescope~\cite{Maggiore:2019uih} and Cosmic Explorer~\cite{Evans:2021gyd}, promise to dramatically increase this number~\cite{Baibhav:2019gxm,Kalogera:2021bya}. Additionally, with the advent of space-based detectors, such as the recently adopted Laser Interferometer Space Antenna (LISA)~\cite{Colpi:2024xhw} or TianQin~\cite{TianQin:2020hid,Li:2024rnk} and Taiji~\cite{Gong:2021any}, the exploration of GWs in the milliHertz regime, will unlock a new class of sources. Among the most exciting sources for fundamental physics, astrophysics and cosmology are intermediate and extreme mass ratio inspirals (I/EMRIs)~\cite{LISA:2022yao,LISACosmologyWorkingGroup:2022jok,LISA:2022kgy}. These systems consist of a (super)massive BH ($\gtrsim 10^{4}M_{\odot}$) orbited by a smaller, typically stellar-mass object. Due to the disparity of masses, the smaller body can spend years in the sensitivity band of detectors, tracing out a complex orbit that requires precise waveform models~\cite{LISAConsortiumWaveformWorkingGroup:2023arg,Barack:2018yvs,Wardell:2021fyy,Chua:2020stf,Katz:2021yft,Hughes:2021exa}. This slow inspiral allows the smaller object to interact {\it dynamically} with any matter configuration surrounding the primary, encoding these interactions into the emitted GWs. As such, I/EMRIs offer a unique opportunity to probe {\it the environments} surrounding BHs~\cite{LISA:2022kgy,Colpi:2024xhw}.
Supermassive BHs are expected to reside at the center of most galaxies, where the existence of environments is ubiquitous. Studies on accretion disks~\cite{1973A&A....24..337S,Novikov:1973kta,2002ApJ...565.1257T,Barausse:2007dy,Abramowicz:2011xu,Yunes:2011ws,Kocsis:2011dr,Barausse:2014tra,Derdzinski:2018qzv,Duffell:2019uuk,Derdzinski:2020wlw,Pan:2021oob,Zwick:2021dlg,Cole:2022yzw,Speri:2022upm,Garg:2022nko,Tiede:2023cje,Zwick:2024yzh,Garg:2024oeu,Garg:2024yrs,Garg:2024zku,Khalvati:2024tzz,Duque:2024mfw}, active galactic nuclei~\cite{Tagawa:2019osr,Tagawa:2020qll,Pan:2021ksp,Derdzinski:2022ltb,Morton:2023wxg}, and dark matter structures~\cite{Eda:2013gg,Macedo:2013qea,Eda:2013gg,Eda:2014kra,Barausse:2014tra,Yue:2018vtk,Hannuksela:2019vip,Kavanagh:2020cfn,Coogan:2021uqv,Dai:2021olt,Cardoso:2021wlq,Coogan:2021uqv,Cole:2022yzw,Cardoso:2022whc,Speeney:2022ryg,Becker:2022wlo,Berezhiani:2023vlo,Karydas:2024fcn,Kavanagh:2024lgq,Bertone:2024wbn}---potentially consisting of ultralight particles~\cite{Ferreira:2017pth,Traykova:2021dua,Kim:2022mdj,Vicente:2022ivh,Bamber:2022pbs,Buehler:2022tmr,Aurrekoetxea:2023jwk,Traykova:2023qyv,Aurrekoetxea:2023jwk,Wang:2024cej,Aurrekoetxea:2024cqd,Dyson:2024qrq} or bosonic clouds~\cite{Zhang:2018kib,Baumann:2018vus,Zhang:2019eid,Baumann:2019ztm,Baumann:2021fkf,Takahashi:2021eso,Cole:2022yzw,Baumann:2022pkl,Takahashi:2023flk,Tomaselli:2023ysb,Brito:2023pyl,Duque:2023seg,Tomaselli:2024bdd,Tomaselli:2024dbw,Boskovic:2024fga,Khalvati:2024tzz}---suggest that GW astronomy is {\it affected} by these environments and could even inform us on the specifics of the matter distribution. In this way, GWs can provide insights into the nature of dark matter or the existence of new fundamental fields. Furthermore, incorporating environmental effects into waveform modeling could not only be essential to actually detect a signal, but also avoid biases in parameter estimation or systematics in tests of general relativity~\cite{Cole:2022yzw,Zwick:2022dih,Garg:2024qxq,Khalvati:2024tzz}. Most previous studies employ Newtonian approximations, even though I/EMRIs probe the strong-field regime of gravity. While some relativistic studies exist, they restrict to spherical symmetry~\cite{Brito:2023pyl,Duque:2023seg,Khalvati:2024tzz} or linear motion~\cite{2007MNRAS.382..826B,Traykova:2021dua,Vicente:2022ivh,Traykova:2023qyv,Wang:2024cej,Dyson:2024qrq}---approaches that are not sufficiently accurate given the precision of future detectors. 
In this work, we present the first self-consistent, fully relativistic calculation of the perturbation on a BH environment induced by an inspiralling secondary in the Kerr geometry. Our formalism is general and can be applied to any non-vacuum spacetime where the primary BH geometry dominates. As a proof of concept, we consider bosonic clouds formed via superradiance, and study the response of the scalar field due to the perturbation of the secondary. We find that the perturbed scalar field develops a distinct spiraling wake trailing behind the secondary. Moreover, the fluxes emitted to infinity and through the horizon show significant deviations---with relative differences of tens of percent compared to the Schwarzschild case, even at relatively large binary separations. These findings emphasize the need to not only model environments relativistically, but also to relax the assumption of spherical symmetry and use the Kerr geometry as a background. Failing to do so could result in serious biases in data analysis with future detectors.
Throughout this work, we use natural units $G = c = \hbar = 1$, unless otherwise stated and we work with a mostly plus metric signature.
%

\noindent{\bf{\em Field Equations with Environments.}} 
We focus on astrophysical systems whose geometry is dominated by the Kerr spacetime, with perturbations arising from the (small) binary companion and surrounding matter. We will first outline our perturbation scheme for generic matter fields.
The action for generic matter fields minimally coupled to gravity in the presence of a perturber with mass $m_{\rm p}$ is given by
\begin{equation}\label{eq:action_generic}
\begin{aligned}
    S \!=\!\int\!\!\mathrm{d}^4 x \sqrt{-\mathbf{g}} \left(\!\frac{\bf R}{16 \pi G} + \mathcal{L}^{\rm env}[ {\bf \Psi}]\!\right)\!- m_{\rm p} \!\int\!\! \mathrm{d}\tau \sqrt{-\mathbf{g}_{\mu\nu}\mathbf{u}^\mu\mathbf{u}^\nu}&\, ,
\end{aligned}
\end{equation}
where $\mathcal{L}^{\rm env}$ is the Lagrangian of the minimally coupled matter field ${\bf \Psi}$ and the action of the point particle encodes the curvature of the secondary BH using the ``skeletonized'' source approach~\cite{Mathisson:1937zz,Dixon:2015vxa}. Here $\mathbf{u^\mu}$ denotes the four-velocity of the secondary on some effective, regularized metric. Bold quantities denote full nonlinear terms.
By varying the action with respect to the matter field and metric, we obtain the field equations:
\begin{align}
    \mathcal{Q} [{\bf \Psi},{\bf g} ]&=0\label{eq:GenericEquationsOfMotionPsi_matter}\,, \\
   G_{\mu \nu}[ {\bf g}]&=  T^{\rm env}_{\mu\nu}[{\bf \Psi},{\bf g} ] +  T^{\rm p}_{\mu\nu}[\gamma] \label{eq:GenericEquationsOfMotionPsi_gravity}\,,
\end{align}
where $\gamma$ is the world-line of the secondary and $\mathcal{Q}$ is a generic nonlinear operator involving ${\bf \Psi}$ and ${\bf g}$.
At leading order, the surrounding matter (henceforth, ``the environment'') is treated as a stationary solution on top of the Kerr geometry and is characterized by a total mass $M^{\rm env}$ and typical length $L^{\rm env}$. The stress-energy tensor of the matter field scales with the energy density as
\begin{equation}
\begin{aligned}
   T^{\rm env}_{\mu\nu} \sim \rho^{\rm env}_0 \sim \frac{M^{\rm env}}{(L^{\rm env})^3}\,.
\end{aligned}
\end{equation}
The natural expansion for the matter field then follows the characteristic density ratio, i.e.,
\begin{equation}\label{eq:EmviromentalExpansionParameter}
\begin{aligned}
   \epsilon^n = \frac{\rho_0^{\rm env}}{\rho_0^{\rm BH}} = \frac{M^{\rm env}} {(L^{\rm env})^3}\frac{L^3}{M} = \eta \left(\frac{L}{L^{\rm env}}\right)^3\,,
\end{aligned}
\end{equation}
where $M$ and $L$ denote the mass and length scale of the primary (Kerr) BH, respectively, and $\eta  \equiv M^{\rm env}/M$. The exponent $n$ corresponds to the leading-order power of ${\bf \Psi}$ in the matter Lagrangian. For instance, $\mathcal{L}^{\rm env} = |\partial{\bf \Psi}|^2|{\bf \Psi}|^2+ |{\bf \Psi}|^6$ implies $n=4$. When $\epsilon\ll 1$, the matter field is then naturally expanded as ${\bf \Psi} = \epsilon \psi + \cdots$ where $\psi$ satisfies the matter field equations on Kerr~\eqref{eq:GenericEquationsOfMotionPsi_matter}.
We define the mass ratio between the primary and secondary BH as $q \equiv m_{\rm p}/M$. Finally, besides $q\ll1$ and  $\epsilon\ll1$, we do not require any scaling relation between $q$ and $\epsilon$:~they act as independent perturbative parameters. To track this dual expansion, we label quantities of some perturbative order $S^{(n,m)} $ as being associated with a perturbative coefficient $\sim \mathcal{O}(\epsilon^n q^m)$.
In this framework, the gravitational and matter fields are expanded as follows:
\begin{align}
    \mathbf{g}_{\mu\nu}&= g_{\mu\nu}+\epsilon^n h^{(n,0)}_{\mu\nu}+q h^{(0,1)}_{\mu\nu}+ \epsilon^n  q  h^{(n,1)}_{\mu\nu}\label{eq:expansion_metric}\\ &+ \epsilon^n q^2   h^{(n,2)}_{\mu\nu}+  q^2  h^{(0,2)}_{\mu\nu}+\cdots\,,\nonumber\\
    \mathbf{\Psi} &= \epsilon \psi^{(1,0)} +\epsilon  q  \psi^{(1,1)}+\epsilon  q^2 \psi^{(1,2)}+\cdots\label{eq:expansion_field}\, ,
\end{align}
where $h^{(0,1)}_{\mu\nu}$ is the metric perturbation arising from a vacuum point-particle source. From the equations of motion, we can then generically expand the nonlinear operators as
\begin{equation}
\begin{aligned}\label{eq:Gexpanded}
    G_{a b }[g_{\mu \nu } + h_{\mu \nu} ] &= G_{ab}[ g_{\mu\nu}] + \delta G_{a b } [h_{\mu \nu }]\\& + \delta^2 G_{a b } [h_{\mu \nu },h_{\mu \nu }] + \ldots\,,
\end{aligned}
\end{equation}
where we defined:
\begin{equation}
     \delta^m G_{ab} [h_{\mu \nu }]  = \frac{1}{m!} \frac{\mathrm{d}^m}{\mathrm{d}\lambda^m} G[g_{\mu\nu}+ \lambda h_{\mu\nu}]|_{\lambda=0}\,.
\end{equation}
Similarly, we expand $\mathcal{Q}$ as
\begin{equation}\label{eq:Qexpanded}
\begin{aligned}
    \mathcal{Q}[\psi+\tilde{\psi},\ &g_{\mu \nu }  +h_{\mu \nu} ] = 
    \mathcal{Q}[\psi, g_{\mu\nu}]+\delta^{(1,0)}\mathcal{Q}[\tilde{\psi},g_{\mu \nu }]  \\
    &\;\;\; + \delta^{(1,1)}\mathcal{Q}[\tilde{\psi},h_{\mu \nu }]+ \delta^{(2,1)}\mathcal{Q}[\tilde{\psi},\tilde{\psi},h_{\mu \nu }] \\
    & \;\;\;+\delta^{(1,2)}\mathcal{Q}[\tilde{\psi},h_{\mu \nu },h_{\mu \nu }] + \ldots\,,
\end{aligned}
\end{equation}
where
\begin{equation}
\begin{aligned}
     \delta^{(n,m)} \mathcal{Q}[\tilde{\psi}, h_{\mu \nu }] & = \frac{1}{n! m!} \frac{\mathrm{d}^{n+m}}{\mathrm{d}\kappa^n \mathrm{d}\lambda^m} \\&\left\{\mathcal{Q}[\psi+ \kappa \tilde{\psi}, g_{\mu\nu}+ \lambda h_{\mu\nu}]\right\}|_{\lambda=0, \kappa=0}\,.
\end{aligned}
\end{equation}
Substituting the perturbed fields~\eqref{eq:expansion_metric}--\eqref{eq:expansion_field} into the expansions of Eqs.~\eqref{eq:GenericEquationsOfMotionPsi_matter}--\eqref{eq:GenericEquationsOfMotionPsi_gravity}, the $\mathcal{O}(\epsilon q)$ perturbation to the environment is found to satisfy:
\begin{equation}\label{eq:dynamicalperubation}
  \delta^{(1,0)}\mathcal{Q}[\psi^{(1,1)}, g_{\mu \nu }] = -  \delta^{(1,1)}\mathcal{Q}[\psi^{(1,0)},h^{(0,1)}_{\mu \nu }]\,,
\end{equation}
which represents the leading-order dynamical perturbation from the secondary BH to the environment.

\noindent{\bf{\em Scalar Fields.}}\label{app:boson_clouds} 
The framework outlined above is general and can be applied to any non-vacuum spacetime with a stress-energy tensor $T^{\rm env}_{\mu \nu}$ satisfying known equations of motion, as long as the perturbations to the environment remain in the linear regime. We will now apply it to the scenario of a massive scalar field around a Kerr BH. 
Through the process of {\it superradiance}~\cite{ZelDovich1971,ZelDovich1972,Starobinsky:1973aij,Brito:2015oca}, massive bosonic fields can extract rotational energy from BHs, forming dense, macroscopic structures known as {\it boson clouds}. This process is most efficient when the Compton wavelength of the field is comparable to the gravitational radius of the BH, i.e., when the ``gravitational coupling'' $\alpha \equiv \mu M \sim \mathcal{O}(0.1)$ (with $\mu$ the boson mass). These clouds can grow on astrophysical timescales~\cite{Detweiler:1980uk,Baumann:2019eav}, reaching a quasi-stationary state known as the {\it superradiance threshold}, where the cloud has a characteristic frequency $\omega_{\rm c}$ and the BH spin parameter is $a/M \approx 4m_{\rm b}\alpha/(m_{\rm b}^2+4\alpha^2)$~\cite{Brito:2014wla,East:2017ovw}. Here, $m_{\rm b}$ is the azimuthal angular momentum of the scalar field. Throughout this work, we assume the cloud to be at this threshold, while residing in its dominant, dipolar ground state $\ket{n_{\rm b}\ell_{\rm b} m_{\rm b}} = \ket{211}$.
The cloud-BH system, or {\it gravitational atom}, has gained significant attention as a target for future GW detectors. The reason is threefold:~(i) the superradiance process is purely gravitational and does not require a pre-existing abundance of the scalar field; (ii) for astrophysical BHs, clouds form for bosons in the mass range $\mathcal{O}(10^{-20}-10^{-10})\,\mathrm{eV}/c^2$. Such ultralight particles are proposed solutions to the strong CP problem~\cite{Weinberg:1977ma,Wilczek:1977pj,Peccei:1977hh} and are plausible dark matter candidates~\cite{Bergstrom:2009ib,Marsh:2015xka,Hui:2016ltb,Ferreira:2020fam}; (iii) studies on the Newtonian level have revealed a rich phenomenology when gravitational atoms are part of a binary system~\cite{Zhang:2018kib,Baumann:2018vus,Zhang:2019eid,Baumann:2019ztm,Baumann:2021fkf,Cole:2022yzw,Baumann:2022pkl,Tomaselli:2023ysb,Tomaselli:2024bdd,Boskovic:2024fga,Tomaselli:2024dbw}, making them interesting candidates for testing fundamental physics with GWs.
We consider a gravitational atom perturbed by a point-particle on an equatorial, circular orbit in the Kerr geometry. The matter Lagrangian for a massive scalar is given by
\begin{equation}\label{eq:action}
\begin{aligned}
\mathcal{L}^{\rm env}[{\bf \Phi}] =  \nabla_\nu {\bf \Phi} \nabla^\nu {\bf \Phi}^* - \mu^2 |{\bf \Phi}|^2\,,
\end{aligned}
\end{equation}
yielding $n=2$ and the equation of motion~\eqref{eq:GenericEquationsOfMotionPsi_matter}:
\begin{equation}
   \mathcal{Q}[{\bf \Phi}, {\bf g}_{\mu\nu}]=\frac{{1}}{\sqrt{- {\bf g}}} \partial_{\mu} (\sqrt{- {\bf g}} \partial^\mu {\bf \Phi}) - \mu^2 {\Phi}\,. 
\end{equation}
Following the previous section, we expand the scalar field as
\begin{equation}\label{eq:expansion_field2}
    \mathbf{\Phi} = \epsilon \phi^{(1,0)} + \epsilon q \phi^{(1,1)}+\epsilon  q^2 \phi^{(1,2)}+\cdots\,.
\end{equation}
We take the background solution $\epsilon \phi^{(1,0)}$ to be that of a superradiant cloud, whose normalization is set such that its total mass is given by $M_{\rm c}$. The characteristic length scale of the cloud is given by its Bohr radius $L^{\rm env} = (\mu \alpha)^{-1} = \alpha^{-2}M$, yielding, $\epsilon = \alpha^3 \sqrt{M_{\rm c}/M}$~\eqref{eq:EmviromentalExpansionParameter} and fixing our perturbative parameter. The expansions of the scalar equation of motion at leading dynamical order $\sim \mathcal{O}(\epsilon q) $ then leads to the expressions:
\begin{align}
 \delta^{(1,0)}\mathcal{Q}[\phi^{(1,1)},g_{\mu \nu }] & =  \frac{ 1}{\sqrt{- g}} \partial_{\mu} (\sqrt{- g} \partial^\mu \phi^{(1,1)}) - \mu^2 \phi^{(1,1)} \nonumber\\& =\left( \Box- \mu^2\right) \phi^{(1,1)}\,, \label{eq:KG_pre}\\
 \delta^{(1,1)}\mathcal{Q}[\phi^{(1,0)},h^{(0,1)}_{\mu \nu }]  &= -
 h^{\mu\nu}_{(0,1)}\nabla_\mu \nabla_\nu \phi^{(1,0)}\label{eq:operators2}\\
 &\mkern-150mu-\frac{h^{(0,1)}}{2} \delta^{(1,0)}\mathcal{Q}[\phi^{(1,0)},g_{\mu \nu }] - \left( \nabla_\mu \bar{h}^{\mu\nu}_{(0,1)}\right) \left( \nabla_\nu \phi^{(1,0)} \right)\,.\nonumber
\end{align}
Here, $\bar{h}_{\mu\nu}$ is the trace-reversed metric perturbation and from now on, all covariant derivatives and d'Alembertian operators are defined with respect to the background metric. As $\phi^{(1,0)}$ is a test-field solution on Kerr, the second term on the right-hand side of Eq.~\eqref{eq:operators2} vanishes. Moreover, as we solve for the $\mathcal{O}(q)$ metric perturbation in Lorenz gauge, i.e., $\nabla_\mu \bar{h}^{\mu\nu}_{(0,1)} = 0$ (see~\cite{Dolan:2023enf,Dolan:2021ijg} and the Supplemental Material (SM)), the final term in Eq.~\eqref{eq:operators2} also vanishes, resulting in the expression:
\begin{equation}\label{eq:rhs_lead}
\begin{aligned}
\delta^{(1,1)}\mathcal{Q}[\phi^{(1,0)},h^{(0,1)}_{\mu \nu }]  = -h_{(0,1)}^{\mu\nu}\nabla_\mu \nabla_\nu \phi^{(1,0)}\,. 
\end{aligned}
\end{equation}
Thus, from Eqs.~\eqref{eq:dynamicalperubation},~\eqref{eq:KG_pre} and~\eqref{eq:rhs_lead}, the leading-order {\it dynamical} perturbation from the secondary to the scalar field is governed by
\begin{equation}\label{eq:EOM_sourced}
\begin{aligned}
\left(\Box - \mu^2\right) \phi^{(1,1)} =  h_{(0,1)}^{\mu\nu}\nabla_\mu \nabla_\nu \phi^{(1,0)}\,.
\end{aligned}
\end{equation}
The specialization to Lorenz gauge results in a source that diverges as $1/|\vec{r} - \vec{r} _{\rm p}|$ at the position of the secondary. At the level of spheroidal harmonics, this leads to a source that is $C^0$ continuous, which makes it particularly well-suited for producing extended solutions across the entire domain.
\begin{figure}
    \centering
    \includegraphics[width=\linewidth]{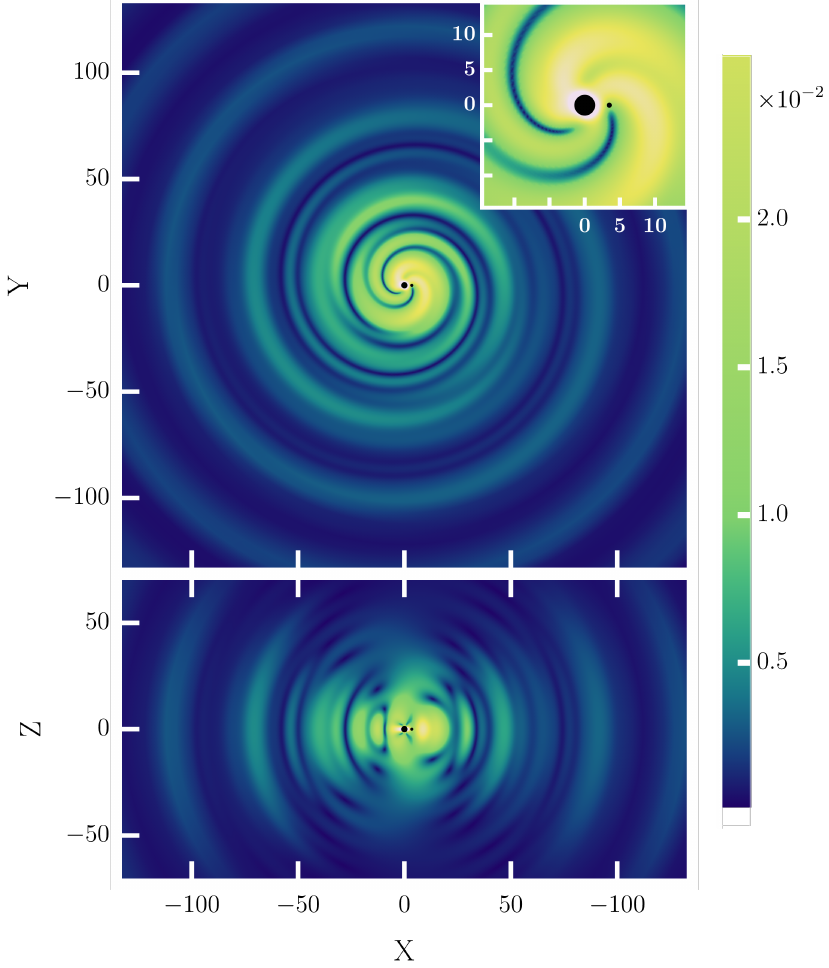}
    \caption{We show the absolute value of the perturbed scalar field $|\phi^{(1,1)}|$ for $\ell\geq 2$, taking $\alpha = 0.3$, $a = 0.88M$ and $r_{\rm p} = 3.5M$. In the top panel, we show an equatorial slice of the field solution, in which the $\hat{Z}$--axis is aligned with the BH spin. In the bottom panel, we show an azimuthal slice of the field, where the secondary moves ``into the plane.''}
    \label{fig:Wake_Profile}
\end{figure}
%

\noindent{\bf{\em Scalar field wake.}} 
Figure~\ref{fig:Wake_Profile} shows the wake profile of the matter field due to a secondary perturber on a prograde equatorial circular orbit. The system parameters are chosen deep in the relativistic regime, with orbital radius $r_{\rm p}=3.5M$, $a=0.88M$ and $\alpha = 0.3$ (corresponding to a cloud whose density peaks at $r\sim 20M$). We solve Eq.~\eqref{eq:EOM_sourced} by decomposing in a spheroidal harmonic basis and find a full solution that constitutes a rich wake structure in the azimuthal and equatorial planes. As the secondary scatters matter through the transfer of angular momentum, a {\it low-density} trail is formed and a spiraling outwash causes matter flux to infinity. 
We also analyze configurations at larger separations (see SM), where certain modes transition from radiative to bound configurations. At these radii, we observe complex changes in the cloud morphology, including configurations where a low-density region forms in front of the secondary, with a {\it high-density} region in its wake. As we always stay in a regime where $\Omega_{\rm p}< \omega_{\rm c}$, this contrasts the picture presented in studies on linear motion of BHs in homogeneous media~\cite{Ostriker_1999, 2007MNRAS.382..826B,Traykova:2021dua,Vicente:2022ivh, Traykova:2023qyv,Wang:2024cej,Dyson:2024qrq}, where a trail of over-density is predicted to form in front of the secondary due to its relative velocity with respect to the background. Consequently, applying results from such studies
to the binary case would yield incorrect conclusions.
%

\noindent{\bf{\em Radiative Energy Loss.} }
As the secondary orbits the central BH, its perturbation induces a transfer of energy and angular momentum to the scalar field and into GWs. It is clear from Eqs.~\eqref{eq:expansion_metric}--\eqref{eq:expansion_field} that many dissipative and conservative effects will drive this change in orbital energy.
\begin{figure*}[t!]
    \centering
    \includegraphics[width=\linewidth]{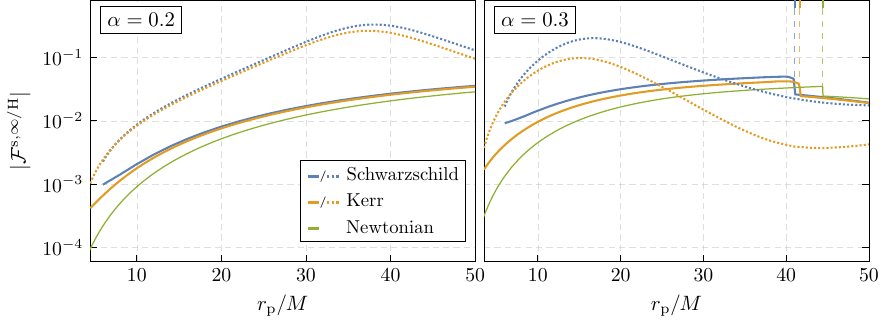}
    \caption{We show the total flux to infinity (solid lines) and through the horizon (dotted lines) considering a prograde orbit and $\alpha = 0.2$ (\emph{left panel}) or $\alpha = 0.3$ (\emph{right panel}). Note that the horizon fluxes are negative on the entire radial domain. The sharp features in the infinity flux in the right panel, computed using Eq.~\eqref{eq:discontinuities}, are marked by vertical dashed lines. Note the Schwarzschild results stop at the innermost stable circular orbit (ISCO) ($r_{\rm p} = 6M$). We sum up to $\ell = 6$ (5) for the infinity (horizon) fluxes.}
    \label{fig:Fluxes_Inf_Hor}
\end{figure*}
For small mass ratios $q$, the secondary evolves adiabatically, meaning that the energy dissipated over one orbit is much smaller than the total orbital energy. Consequently, the evolution of the secondary can be constructed with a sequence of geodesics~\cite{Barack:2018yvs,Wardell:2021fyy,Chua:2020stf,Katz:2021yft,Hughes:2021exa}. As a first step towards determining how the secondary moves from one geodesic to the next, we can consider the leading-order flux balance of the system:
\begin{equation}\label{eq:fluxbalance_main}
    \dot{E}_{\rm orb} + \dot{M}_{\rm c} = -\dot{E}^{\scalebox{0.7}{$\mathrm{GW}$},\infty}-\dot{E}^{\scalebox{0.7}{$\mathrm{GW}$}, \mathrm{H}}-\dot{E}^{\Phi,\,\infty}-\dot{E}^{\Phi,\mathrm{H}}\,.
\end{equation}
Defining the ``scalar flux,'' of the environment as $\mathcal{F}^{\mathrm{s}, \infty/\mathrm{H}} \equiv \epsilon^{-2} q^{-2}\left( \dot{E}^{\Phi,\,\infty/\mathrm{H}}+ \dot{M}_{\rm c}^{\infty/\mathrm{H}} \right)$, we can then calculate its emission to infinity and through the horizon, and apply Eq.~\eqref{eq:fluxbalance_main} to find $\dot{E}_{\rm orb}$. Importantly, this equation ignores conservative contributions from $h^{(2,0)}$ and $h^{(2,1)}$, which arise at the same order and should be calculated in future work. In Fig.~\ref{fig:Fluxes_Inf_Hor}, we show the scalar fluxes for $\alpha = 0.2$ and $\alpha = 0.3$, corresponding to highly spinning BHs with $a = 0.69M$ and $a = 0.88M$, respectively. For comparison, we also include the fluxes obtained in the Newtonian case~\cite{Baumann:2021fkf,Baumann:2022pkl,Tomaselli:2023ysb};~further details can be found in SM. The main features are:
(i)~as the orbital separation between the binary components grows, effects related to spin of the primary become small for the infinity flux, and Schwarzschild and Kerr become similar. Similarly, the green line, belonging to the Newtonian regime, converges toward the relativistic cases at large radii. We find the relative differences between Kerr and Schwarzschild is tens of percent, reaching 50\% near the ISCO (see SM).
(ii)~the horizon flux exceeds the infinity flux across most of the shown radial domain. It is dominated by the $(\ell,m) = (0,0)$ mode and always negative, indicating that the binary's orbit gains energy. This is due to a resonance between bound states of the cloud:~the initial $\ket{211}$ state resonates with $\ket{100}$, which has lower energy and angular momentum. This surplus is fed back into the orbit, potentially giving rise to a {\it floating orbit}~\cite{Baumann:2019ztm,Tomaselli:2024dbw,Tomaselli:2024bdd}, where the binary's evolution is slowed down or even stalled for a period of time. For $\alpha = 0.3$, the $(\ell,m) = (2,2)$ mode becomes significant around $r_{\rm p} \sim 50M$, nearly overtaking the $(0,0)$ contribution. Unlike the $(0,0)$ mode, it produces a positive horizon flux, inducing a {\it sinking orbit}:~a period of accelerated inspiral. Such resonances are key observables for probing the cloud's properties, yet they are also efficient at depleting the cloud itself~\cite{Tomaselli:2024dbw,Tomaselli:2024bdd}.
(iii) the right panel reveals a sharp feature in the flux to infinity, consistent with earlier studies~\cite{Baumann:2021fkf,Baumann:2022pkl,Tomaselli:2023ysb,Brito:2023pyl}. These arise when a new mode starts contributing to the flux, specifically, occurring when
\begin{equation}\label{eq:discontinuities}
 r^{*,m}_{\rm p} = \left(\frac{m-m_{\rm b}}{\mu-\text{Re}[\omega_{\rm c}]}-a\right)^{2/3}M\,.
\end{equation}
Using Eq.~\eqref{eq:discontinuities} and calculating $\omega_{\rm c}$ with Leaver's method~\cite{Leaver:1985ax,Dolan:2007mj}, we find that for $\alpha = 0.3$, $r^{*,2}_{\rm p}/M = 41.01, 41.66, 44.44$ in the Schwarzschild, Kerr and Newtonian case, respectively, in precise agreement with Fig.~\ref{fig:Fluxes_Inf_Hor} (vertical dashed lines). Notably, sharp features are absent in the $\alpha = 0.2$ case as they occur at larger radii (e.g.~$r^{*,2}_{\rm p} \sim 100M$). The physical origin of these features lies in the long-range nature of the gravitational potential, as detailed in App.~D of~\cite{Baumann:2021fkf}. Finally, close to $r^{*,m}_{\rm p}$, the wavelength of the modes becomes extremely large, which requires the flux to be extracted far out. The small dip in the flux preceding the feature is thus merely a numerical artifact.
(iv) consistent with previous studies~\cite{Baumann:2021fkf,Baumann:2022pkl,Brito:2023pyl,Duque:2023seg}, we observe that scalar fluxes tend to dominate over gravitational fluxes during the early inspiral stage. As the gravitational and scalar fluxes rely on independent perturbative parameters, a general comparison with the results in Fig.~\ref{fig:Fluxes_Inf_Hor} should not be made. However, an example case for a given $q$ and $\epsilon$ is provided in the SM.
Our results for the scalar flux in Schwarzschild are not in full agreement with previous work~\cite{Brito:2023pyl}, which used a different gauge. We find a discrepancy of up to $\sim 10\%$ (see SM). A plausible source of this discrepancy lies in the nature of the background solution in Schwarzchild, which includes an exponentially decaying term $\sim e^{-\mathrm{Im}[\omega_{\rm c}] t}$, where $\mathrm{Im}[\omega_{\rm c} M]$ is $\sim 1-10\%$ the value of the unnormalized flux. This decaying behavior makes the background ill-suited as a stationary state to perturb around in frequency domain. To address this, we follow~\cite{Brito:2023pyl} and set $\mathrm{Im}[\omega_{\rm c} M] =0 $ by hand, which means the background solution is not an exact solution of the homogeneous Klein-Gordon equation. This approximation introduces gauge dependence in the asymptotic values of the perturbed scalar field and violates the conservation laws underpinning Eq.~\eqref{eq:fluxbalance_main}. This issue does not arise in Kerr where a stationary background solution can be found.
%

\noindent{\bf{\em Discussion.}} 
In this {\it Letter}, we develop a framework to study how generic BH environments are perturbed in EMRIs. As an example case, we apply it to boson clouds and, for the first time, self-consistently compute the perturbation of an EMRI to the environment in the Kerr geometry. We have demonstrated the importance of performing these calculations in Kerr, by comparing them with fluxes in Schwarzschild. For less relativistic environments ($\alpha = 0.2$), where the density peaks at $\sim 50M$, we find relative differences of around 10\%, increasing to $50\%$ near the ISCO. In more relativistic environments ($\alpha = 0.3$), these differences are even more significant, attaining $30-100\%$ throughout the region where EMRIs are expected to enter the LISA band.
Additionally, we solved for the field perturbations across the entire domain, revealing a rich wake structure induced by the secondary. Our results demonstrate how the morphology of the environment changes with the position of the secondary, emphasizing the intricate and rich dynamics of these systems, which are linked to striking observational signatures with future GW detectors~\cite{Baumann:2018vus,Baumann:2019ztm,Baumann:2021fkf,Cole:2022yzw,Baumann:2022pkl,Tomaselli:2023ysb,Brito:2023pyl,Duque:2023seg,Tomaselli:2024bdd,Boskovic:2024fga,Tomaselli:2024dbw,Khalvati:2024tzz}. These results raise important questions about existing studies that use linear motion of BHs in a homogeneous medium as a proxy for dynamical friction in a binary inspiral (e.g.~\cite{Kavanagh_2020,Coogan:2021uqv}), showing that such approximations, or using Schwarzschild as a background is inadequate and will lead to significant errors. Instead, the correct approach in perturbation theory is to use the framework we have developed.
We expect this work to serve as the starting point for self-consistent modeling of EMRIs and environments in Kerr. There are several directions that warrant further exploration in the future. For instance, applying our framework to the Navier-Stokes system would provide a crucial step toward understanding EMRI dynamics in accretion disks in the fully relativistic regime. Another key challenge still lies in calculating the conservative and dissipative effects of all field perturbations, and we have yet to explore the slow-time contributions inherent to these systems. A two-timescale analysis will be necessary to understand how all of these contributions affect the binaries' orbital parameters.
%

\noindent{\bf{\em Acknowledgements.}} We are grateful to Vitor Cardoso for comments on the final draft of this manuscript. C.D., M.v.d.M. and T.S. are supported by the VILLUM Foundation (grant no. VIL37766), the Danish Research Foundation (grant no. DNRF162), and the European Union’s H2020 ERC Advanced Grant ``Black holes: gravitational engines of discovery'' grant agreement no. Gravitas–101052587. 
R.B. acknowledges financial support provided by FCT – Fundação para a Ciência e a Tecnologia, I.P., under the Scientific Employment Stimulus -- Individual Call -- Grant No. \href{https://doi.org/10.54499/2020.00470.CEECIND/CP1587/CT0010}{2020.00470.CEECIND}, the Project No. \href{https://doi.org/10.54499/2022.01324.PTDC}{2022.01324.PTDC} and the Project ``GravNewFields'' funded under the ERC-Portugal program. S.D.~acknowledges financial support from the Science and Technology Facilities Council (STFC) under Grant No.~ST/X000621/1 and Grant No.~ST/W006294/1. This work makes use of the Black Hole Perturbation Toolkit. The Tycho supercomputer hosted at the SCIENCE HPC center at the University of Copenhagen was used for supporting this work. 

\bibliography{ref}
\newpage
\renewcommand{\thesubsection}{{S.\arabic{subsection}}}
\setcounter{section}{0}
\section*{Supplemental material}
\subsection{Lorenz Gauge Metric Perturbations.}\label{app:Lorenz} 
Diffeomorphism invariance of General Relativity becomes \emph{gauge freedom} in perturbation theory. Under a change of coordinates $x^\mu \rightarrow x^\mu + \varepsilon X^\mu(x)$, the linear perturbation of a tensor $\mathbf{T}$ transforms as $\delta T \rightarrow \delta T - \pounds_X T$, where $\pounds_X T$ is the Lie derivative of $T$ on the background; consequently, the linear metric perturbation transforms as $h_{\mu \nu} \rightarrow h_{\mu \nu} - 2 \nabla_{(\mu} X_{\nu)}$.
For many calculations, it is convenient to exploit gauge freedom to work in Lorenz gauge, defined by
\begin{equation}
\nabla^\mu \bar{h}_{\mu \nu}^{(n,m)} = 0\,, 
\end{equation}
where $\bar{h}_{\mu \nu} = h_{\mu \nu} - \frac{1}{2} g_{\mu \nu} h$ is the trace-reversed metric perturbation, and $h = g^{\mu \nu} h_{\mu \nu}$ is the trace (indices $(n,m)$ omitted for clarity). The covariant derivative $\nabla_\mu$ is defined on the background spacetime $g_{\mu \nu}$. Then the perturbed Einstein equations become a system of hyperbolic equations, 
\begin{equation}\label{eq:LEE} 
\Box \barh^{(n,m)}_{\mu \nu} + 2 \tensor{R}{^\alpha _\mu ^\beta _\nu} \barh^{(n,m)}_{\alpha \beta}  
= S^{(n,m)}_{\mu\nu}\,,    
\end{equation}
where the source term $S^{(n,m)}_{\mu\nu}$ depends on metric and field perturbations of lower orders.
Recent work~\cite{Dolan:2021ijg, Dolan:2023enf, Wardell:2024yoi} has developed a prescription for constructing the Lorenz-gauge metric perturbation on the Kerr background from scalar variables that satisfy decoupled, separable equations; more specifically, sourced Teukolsky equations of spin-2, spin-1 and spin-0 types \cite{Teukolsky:1973ha}. In this paper we make use the implementation in Ref.~\cite{Dolan:2023enf} to compute the metric perturbation of a pointlike body on a circular equatorial orbit of a rotating BH in Lorenz gauge.
\subsection{Fluxes}\label{app:fluxes}
In addition to GWs, the secondary will induce emission of scalar waves, both to infinity and through the horizon of the primary. These will in turn determine how the secondary evolves. Here, we derive the explicit form of the flux formulae in the case of boson clouds, and compare our results with previous work in Schwarzschild~\cite{Brito:2023pyl}. In this section, we suppress much of the perturbative indexing in the interest of readability. However, the order of most quantities should be clear from context.
\subsubsection{Flux formulae}
The (orbit-averaged) energy fluxes of the perturbed field to infinity and through the horizon can be calculated, respectively, as~\cite{Teukolsky:1973ha,Teukolsky:1974yv} 
\begin{equation}\label{eq:energy_fluxes}
\begin{aligned}
\dot E^{\Phi,\infty} &= -\lim_{r\to +\infty} r^2\int \mathrm{d}\Omega\, T^{\Phi}_{\mu r}\xi_{(t)}^{\mu}\,,\\ 
\dot E^{\Phi,\mathrm{H}} &= \lim_{r\to r_+} 2M r_+ \int \mathrm{d}\Omega\, T^{\Phi}_{\mu\nu}\xi_{(t)}^{\mu}l^{\nu}\,,
\end{aligned}
\end{equation}
where 
\begin{equation}\label{eq:stress}
T^{\Phi}_{\mu\nu}  = \nabla_{(\mu} \Phi \nabla_{\nu)} \Phi^* - \frac{g_{\mu\nu}}{2}  ( \nabla_{\delta}\Phi\nabla^{\delta}\Phi^* + \mu^2 |\Phi|^2 )\, ,
\end{equation}
and $\mathrm{d}\Omega$ is the area element of the $2$--sphere, $l^{\mu}=\partial/\partial t+\Omega_{\rm H}\partial/\partial \varphi$ is a null vector normal to the horizon, with $\Omega_{\rm H}=a/(2Mr_+)$ and $\xi_{(t)}^{\mu}\equiv\partial/\partial t$, $\xi_{(\varphi)}^{\mu}\equiv\partial/\partial\varphi$ are the Killing vectors of the Kerr metric. Analogous expressions for angular momentum fluxes are obtained by swapping $\xi_{(t)}^{\mu} \rightarrow \xi_{(\varphi)}^{\mu}$ in Eq.~\eqref{eq:energy_fluxes}.
We define the mass of the background boson cloud to be given by the volume integral on a spacelike slice of the time component of the stress-energy tensor. In particular, we define
\begin{equation}
    M_{\rm c} = \int_{r_{\rm H}}^{\infty}  \int_{S^2} T_t^t[\phi^{(1,0)},g^{\text{Kerr}}] r^2 \mathrm{d}r\,  \mathrm{d}\Omega\,.
\end{equation}
We decompose $\phi^{(1,1)}$ in a spheroidal harmonics basis, 
\begin{equation}
    \phi^{(1,1)} = \sum_{\ell,m} \phi^{(1,1)}_{\ell m}(r)S_{\ell m}(\theta,\varphi,\gamma_{m_{\rm g}})e^{- i (\Omega_{m_{\rm g}} + \omega_{\rm c}) t}\,,
\end{equation}
where $\gamma_{m_{\rm g}} =  a \sqrt{(\Omega_{m_{\rm g}}+\omega_{\rm c})^2-\mu^2}$ and for circular orbits $\Omega_{m_{\rm g}} = (m- m_{\rm b})\Omega_{\rm p} \equiv m_{\rm g}\Omega_{\rm p}$. The above equations lead to the following expressions for the energy fluxes of individual modes:
\begin{equation}\label{eq:enery_flux_ind}
\begin{aligned}
    \dot E^{\Phi,\infty}_{\ell m} &= \lim_{r\to +\infty} r^2\Big\{2\,|\omega_{\rm c} + \Omega_{m_{\rm g}}|\\& \times\mathrm{Re}\left[\sqrt{\left(\Omega_{m_{\rm g}}+\omega_{\rm c}\right)^{2}-\mu^2}\right]|\phi^{(1,1)}_{\ell m}|^{2} \Big\}\,,\\
    \dot E^{\Phi,\mathrm{H}}_{\ell m} &= \lim_{r\to r_+} 2M r_+\Big\{2\, \left(\omega_{\rm c}+\Omega_{m_{\rm g}}\right)\\ &\times \left(\omega_{\rm c}+\Omega_{m_{\rm g}} - m\Omega_{\rm H}\right)|\phi^{(1,1)}_{\ell m}|^{2}\Big\}\,.
\end{aligned}
\end{equation}
Here, we have set the frequency at the superradiant threshold, i.e., $\omega = \omega_{\rm c}$ and we remind the reader that  at leading-order $\phi^{(1,1)}_{\ell m} \propto 1/r$ when $r\to\infty$. 
Similarly, the angular momentum fluxes are given by
\begin{equation}\label{eq:angmom_flux_ind}
\begin{aligned}
    \dot L^{\Phi,\infty}_{\ell m} &= \lim_{r\to +\infty} r^2\Big\{ 2m s_{m_{\rm g}}\\&\times\mathrm{Re}\left[\sqrt{\left(\Omega_{m_{\rm g}}+\omega_{\rm c}\right)^{2}-\mu^2}\right]|\phi^{(1,1)}_{\ell m}|^{2} \Big\}\,,\\
    \dot L^{\Phi,\mathrm{H}}_{\ell m} &= \lim_{r\to r_+} 2 M r_+\left\{ 2m\left(\omega_{\rm c}+\Omega_{m_{\rm g}} - m\Omega_{\rm H}\right)|\phi^{(1,1)}_{\ell m}|^{2}\right\}\,,
\end{aligned}
\end{equation}
where $s_{m_{\rm g}} \equiv \mathrm{sgn}\left(\omega_{\rm c} + \Omega_{m_{\rm g}}\right)$. While $\dot{E}^{\Phi,\,\infty/\mathrm{H}}$ can be truly associated as a ``scalar flux,'' the scalar perturbations also affect the cloud, which in turn impacts the evolution of the secondary. Assuming an adiabatic evolution and conservation of energy and angular momentum, we thus have:
\begin{equation}\label{eq:fluxbalance2}
\begin{aligned}
    \dot{E}_{\rm orb} + \dot{M}_{\rm c} &= -\dot{E}^{\scalebox{0.7}{$\mathrm{GW}$},\infty}-\dot{E}^{\scalebox{0.7}{$\mathrm{GW}$}, \mathrm{H}}-\dot{E}^{\Phi,\,\infty}-\dot{E}^{\Phi,\mathrm{H}}\,,\\
    \dot{L}_{\rm orb} + \dot{S}_{\rm c} &= -\dot{L}^{\scalebox{0.7}{$\mathrm{GW}$},\infty}-\dot{L}^{\scalebox{0.7}{$\mathrm{GW}$}, \mathrm{H}}-\dot{L}^{\Phi,\infty}-\dot{L}^{\Phi,\mathrm{H}}\,,
\end{aligned}
\end{equation}
where $S_{\rm c}$ is the spin of the cloud. This balance equation thus allows one to evolve orbital parameters due to energy emission from the secondary and/or environment. Importantly, this formula excludes the effects of conservative energy transfer between the orbit and bound states of the cloud. As such, it should be used with some caution until all contributions up to order $\mathcal{O}(\epsilon^2 q^2)$ (e.g., $h^{(2,1)}$) are fully understood. Nevertheless, in the absence of a complete understanding, this equation can be used as a first step towards producing time-domain evolutions and generating relativistic waveforms for EMRIs with environments in Kerr. Some insights into how such conservative transfer might occur in a relativistic setting have been studied in~\cite{Redondo-Yuste:2023snb}.
To compute the rate at which the mass and spin of the cloud changes, we make use of the global $U(1)$ symmetry of the (complex) scalar field, whose conserved current implies the existence of a conserved Noether charge $Q$:
\begin{equation}
    Q = \int_{\Sigma} \mathrm{d}^{3}x\sqrt{-g}\,j^0_{\Phi}\,,
\end{equation}
where $\Sigma$ is a space-like hypersurface and
\begin{equation}\label{KG_current}
    j^{\Phi}_{\mu} = -i\left(\Phi^{*}\partial_\mu\Phi-\Phi\partial_\mu\Phi^{*}\right)\,.
\end{equation}
The mass and spin of the cloud are then related to the cloud's Noether charge, i.e., $M_{\rm c} = \omega_{\rm c} Q$, $S_{\rm c} = m_{\rm b}Q$, respectively. The rate of change of the scalar charge is
\begin{equation}\label{eq:charge_fluxes}
\begin{aligned}
\dot Q^{\Phi,\infty} &= -&&\lim_{r\to +\infty} r^2\int \mathrm{d}\Omega\, j^{\Phi}_{r}\,,\\ 
\dot Q^{\Phi,\mathrm{H}} &= &&\lim_{r\to r_+} 2M r_+ \int \mathrm{d}\Omega\, j^{\Phi}_{\mu}l^{\mu}\,,
\end{aligned}
\end{equation}
leading to:
\begin{equation}\label{eq:Noether_ind}
\begin{aligned}
    \dot Q^{\Phi,\infty}_{\ell m} &= -\lim_{r\to +\infty} r^2\Big\{ 2\,s_{m_{\rm g}}\\&\times \mathrm{Re}\left[\sqrt{\left(\Omega_{m_{\rm g}}+\omega_{\rm c}\right)^{2}-\mu^2}\right]|\phi^{(1,1)}_{\ell m}|^{2} \Big\}\,,\\
    \dot Q^{\Phi,\mathrm{H}}_{\ell m} &= -\lim_{r\to r_+} 2 M r_+\left\{ 2\,\left(\omega_{\rm c}+\Omega_{m_{\rm g}} - m\Omega_{\rm H}\right)|\phi^{(1,1)}_{\ell m}|^{2}\right\}\,.
\end{aligned}
\end{equation}
Through the Noether charge, we can then define the scalar energy and angular momentum ``power'' as
\begin{equation}
\begin{aligned}
    \dot{E}^{\mathrm{s},\infty/\mathrm{H}} &= \dot{E}^{\Phi,\infty/\mathrm{H}} + \omega_{\rm c} \dot{Q}^{\Phi,\infty/\mathrm{H}}\,,\\
    \dot{L}^{\mathrm{s},\infty/\mathrm{H}} &= \dot{L}^{\Phi,\infty/\mathrm{H}} + m_{\rm b} \dot{Q}^{\Phi,\infty/\mathrm{H}}\,.
\end{aligned}
\end{equation}
Plugging in Eqs.~\eqref{eq:enery_flux_ind}, \eqref{eq:angmom_flux_ind} and~\eqref{eq:Noether_ind} in the balance equation~\eqref{eq:fluxbalance2}, we then find the explicit terms shown in Fig.~\ref{fig:Fluxes_Inf_Hor}:
\begin{equation}\label{eq:scalar_flux}
\begin{aligned}
    \dot E^{\mathrm{s},\infty}_{\ell m} &= \lim_{r\to +\infty} r^2\Big\{ 2\,\Omega_{m_{\rm g}}s_{m_{\rm g}}\\&\times\mathrm{Re}\left[\sqrt{\left(\Omega_{m_{\rm g}}+\omega_{\rm c}\right)^{2}-\mu^2}\right]|\phi^{(1,1)}_{\ell m}|^{2} \Big\}\,,\\
    \dot E^{\mathrm{s},\mathrm{H}}_{\ell m} &= \lim_{r\to r_+} 2 M r_+\left\{ 2\Omega_{m_{\rm g}}\left(\omega_{\rm c}+\Omega_{m_{\rm g}} - m\Omega_{\rm H}\right)|\phi^{(1,1)}_{\ell m}|^{2}\right\}\,.
\end{aligned}
\end{equation}
Finally, for the angular momentum, we have
\begin{equation}
\begin{aligned}
    \dot L^{\mathrm{s},\infty}_{\ell m} &= \lim_{r\to +\infty} r^2\Big\{ 2m_{\rm g} s_{m_{\rm g}}\\&\times\mathrm{Re}\left[\sqrt{\left(\Omega_{m_{\rm g}}+\omega_{\rm c}\right)^{2}-\mu^2}\right]|\phi^{(1,1)}_{\ell m}|^{2} \Big\}\,,\\
    \dot L^{\mathrm{s},\mathrm{H}}_{\ell m} &= \lim_{r\to r_+} 2 M r_+\left\{ 2m_{\rm g}\left(\omega_{\rm c}+\Omega_{m_{\rm g}} - m\Omega_{\rm H}\right)|\phi^{(1,1)}_{\ell m}|^{2}\right\}\,,
\end{aligned}
\end{equation}
which satisfy
\begin{equation}\label{eq:energy_angmom}    \dot{E}^{\mathrm{s},\infty/\mathrm{H}} = \Omega_{\rm p} \dot{L}^{\mathrm{s},\infty/\mathrm{H}}\,.
\end{equation}
Consequently, the backreaction from the leading-order scalar fluxes onto the secondary evolves circular orbits into circular orbits, justifying the quasi-circular limit studied in this work. Because of this relation~\eqref{eq:energy_angmom}, it is also sufficient to look at the energy fluxes only, as is done in the main text.

\begin{figure*}
    \centering
    \includegraphics[width=\linewidth]{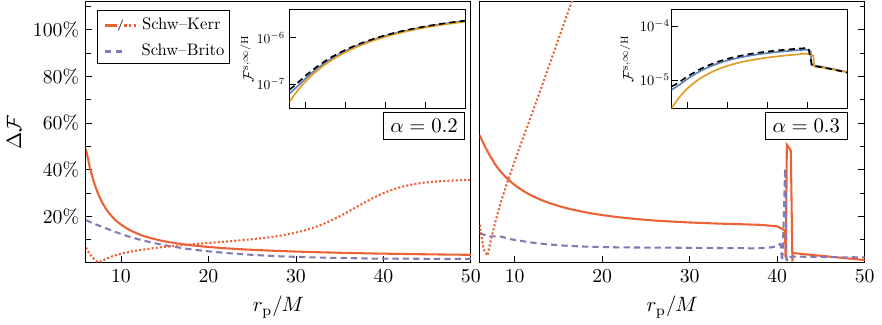}
    \caption{We show the relative error $\Delta \mathcal{F}$ in the total flux to infinity (solid lines) and through the horizon (dotted lines), taking our Schwarzschild data as a reference. We consider a prograde orbit and $\alpha = 0.2$ (\emph{left panel}) or $\alpha = 0.3$ (\emph{right panel}). The comparison is between our Kerr and Schwarzschild (red lines) and between our Schwarzschild and the Schwarzschild from an earlier work~\cite{Brito:2023pyl} (purple line), which used a different gauge. In the inset, we show the actual flux $\mathcal{F}^{\mathrm{s}, \infty}$ in Schwarzschild/Kerr from our data (blue/orange solid) compared to the one from~\cite{Brito:2023pyl} (black dashed). The horizontal axis is the same as in the main plot.}
    \label{fig:RelDiffFluxes_Inf_Hor}
\end{figure*}
\subsubsection{Comparisons}
We shortly detail the derivation of the Newtonian fluxes~\cite{Baumann:2021fkf,Baumann:2022pkl,Tomaselli:2023ysb} shown in the main text (Fig.~\ref{fig:Fluxes_Inf_Hor}) and compare the fluxes between Kerr and Schwarzschild, as well as between our Schwarzschild results and those from an earlier work~\cite{Brito:2023pyl}.
The study of dynamical friction in gravitational atoms in the Newtonian regime has been referred to as {\it ionization}~\cite{Baumann:2021fkf}, due to analogy with atomic physics (see~\cite{Tomaselli:2023ysb} for a thorough comparison between ``classic'' dynamical friction and ionization). Below, we summarize the ionization process and refer to~\cite{Baumann:2021fkf} for details.
In the language of quantum mechanics, ionization describes the transfer of the cloud from its bound state $\ket{n_{\rm b}\ell_{\rm b} m_{\rm b}}$, to any unbound state $\ket{k \ell m}$, where $k$ represents the wavenumber. This process is governed by the coupling strength between these states, defined by the matrix element: 
\begin{equation}
    \eta = \bra{k\ell m}V\ket{n_{\rm b}\ell_{\rm b} m_{\rm b}}\,,
\end{equation}
where $V$ is the gravitational perturbation of the secondary, which is expressed as a multipole expansion of the Newtonian potential. 
The ``ionization power'' is then found by summing over all the unbound states. On circular, equatorial orbits, it is given by 
\begin{equation}
    P_{\rm ion} = \frac{M_{\rm c}}{\mu}\sum_{\ell, m} \Omega_{m_{\rm g}} |\eta(k_*)|^{2}\Theta(k_*^2)\,,
\end{equation}
where $k_* = -\mu \alpha^2/(2n_{\rm b}^2)\pm \Omega_{m_{\rm g}}$ and $\Theta$ is the Heaviside step function. This quantity is equivalent to the energy flux to infinity, $P_{\rm ion} \equiv \dot{E}^{\mathrm{s}, \infty}$.
In Schwarzschild, massive scalar fields still settle on quasi-bound states, even though spin is essential for superradiance to occur. However, these states are decaying ($M\mathrm{Im}[\omega_{\rm c}] < 0$) due to absorption at the horizon, preventing the cloud from achieving a stationary configuration. In the Schwarzschild case, following~\cite{Brito:2023pyl}, we explicitly ignore this, setting $M\mathrm{Im}[\omega_{\rm c}]  = 0$.
In the main text (Fig.~\ref{fig:Fluxes_Inf_Hor}), we compare fluxes in the Newtonian, Schwarzschild and Kerr regimes. Here, we focus on comparing in detail two cases:~Kerr versus Schwarzschild, and ``our'' Schwarzschild results versus those from an earlier work~\cite{Brito:2023pyl} which used a different method and gauge. The relative differences in the fluxes at infinity and at the horizon are shown in Fig.~\ref{fig:RelDiffFluxes_Inf_Hor}, using our Schwarzschild results as the reference.
As expected, at large radii, the Schwarzschild and Kerr flux become similar, with relative differences dropping to a few percent at $r_{\rm p} = 50M$. Closer to the ISCO, the differences become larger, emphasizing the necessity of performing calculations in Kerr and not in Schwarzschild. When comparing our Schwarzschild results to those in~\cite{Brito:2023pyl}, we find relative differences up to $20\%$, due to gauge ambiguities, as discussed in the main text. In addition, mode-by-mode comparisons reveal notable differences between our results and~\cite{Brito:2023pyl}, particularly in the quadrupole.
In the relativistic case, we sum up to $\ell = 6$, ensuring that the flux increment remains below $1\%$ across the considered radial domain. In the Newtonian case instead, the computational cost is much lower, allowing us to easily sum up to $\ell = 10$, which pushes the flux increment below $0.01\%$. At larger radii than we are showing in this work, more modes might be required to obtain accurate results, which poses a computational challenge in the relativistic regime. Therefore, to compute the total flux across a large radial domain, e.g., for waveform modeling in packages like \texttt{FEW}~\cite{Chua:2020stf,Katz:2021yft,Hughes:2021exa}, a smooth interpolation between both approaches might be required.
Importantly, the scalar flux we calculate does not describe the complete flux at $\mathcal{O}(\epsilon^2q^2)$. This is due to the fact that there are two additional gravitational contributions to the energy fluxes which we do not consider. One is due to the expansion of the Einstein operator, $\delta^2G[h^{(2,1)},h^{(0,1)} ]$ which arises as an additional term in the integrand of $\mathcal{F}^{\mathrm{s}, \infty/\mathrm{H}}_{(2,2)}$, which we do not calculate. Additionally, there is a term coming from the $\mathcal{O}(\epsilon^2)$ correction to the orbital frequency ($\Omega_{\rm p}^{(2,0)}$) from the background matter field.
Thus, to calculate the full gauge invariant flux at $\mathcal{O}(\epsilon^2q^2) $ one requires the full expression:
\begin{equation}\label{eq:gaugeinvariantflux}
\mathcal{F}^{ \infty/\mathrm{H}}_{(2,2)} \equiv \mathcal{F}^{\mathrm{s}, \infty/\mathrm{H}}_{(2,2)} + \frac{\mathrm{d}\mathcal{F}^{\scalebox{0.7}{$\mathrm{GW}$}, \infty/\mathrm{H}}_{(0,2)}}{\mathrm{d} \Omega_{\rm p}^{\text{Kerr}}}\Omega_{\rm p}^{(2,0)}\,,
\end{equation}
where $\mathcal{F}^{\mathrm{s}, \infty/\mathrm{H}}_{(2,2)}$ should be corrected to include the term arising from the expansion of the Einstein operator and $\mathcal{F}^{\scalebox{0.7}{$\mathrm{GW}$}, \infty/\mathrm{H}}_{(0,2)}$ is the first order vacuum flux due to the secondary. Moreover, $\Omega_{\rm p}^{\text{Kerr}}$ is the frequency in Kerr for a given orbital radius. The calculation of the $h^{(2,0)}$ and $h^{(2,1)}$ metric perturbations is necessary to calculate these additional terms and is still an open problem.
\subsection{Scalar vs GW flux}\label{app:scalarvsGW}
%
\begin{figure}
    \centering
    \includegraphics[width=\linewidth]{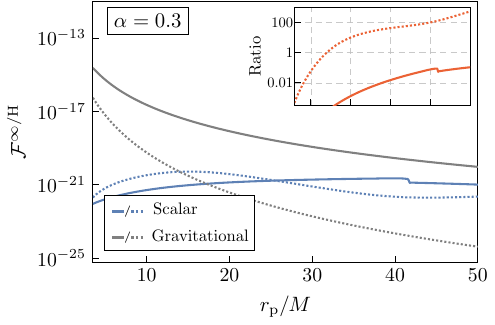}
    \caption{We show the fluxes to infinity and the horizon for the scalar and gravitational case, including the correct perturbative prefactors, considering $q = 10^{-6}$, $\epsilon^2 = 0.1\alpha^3$, $\alpha = 0.3$. In the inset, we show the ratio between the two, i.e., $\mathcal{F}_{\rm s}/\mathcal{F}_{\scalebox{0.6}{$\mathrm{GW}$}}$. The horizontal axis is the same as in the main plot.}
    \label{fig:RatioFluxes_Inf_Hor}
\end{figure}
Previous works have shown that scalar fluxes can dominate over gravitational fluxes at large radii. While we cannot directly compare the two quantities, as they come in at different perturbative orders, i.e., $q$ and $\epsilon$ are independent,  we can consider an example system. Since both quantities scale as $q^2$, this factor cancels out in their ratio. In Fig.~\ref{fig:RatioFluxes_Inf_Hor}, we show a case for which the cloud has obtained a maximum mass $\eta = M_{\rm c}/M = 0.1$~\cite{Brito:2014wla,East:2017ovw,Herdeiro:2021znw}, with a typical EMRI mass ratio $q = 10^{-6}$ and $\alpha = 0.3$. The results show that scalar horizon fluxes quickly dominate over the gravitational horizon fluxes, while the fluxes to infinity will probably overtake at larger radii. 
Finally, we note that the horizon flux is less relevant in spherically symmetric structures that were studied before~\cite{Brito:2023pyl,Duque:2023seg}. A possible reason is that for spherical structures the $(\ell,m)=(0,0)$ mode does not contribute to $\dot E^{\mathrm{s,H}}$, unlike for dipolar clouds. Since there is no angular barrier for $(\ell,m)=(0,0)$ modes, those are more ``easily'' absorbed at the horizon.
%

\subsection{Field Resonances}\label{app:resonances}
\begin{figure}
    \centering
    \includegraphics[width=\linewidth]{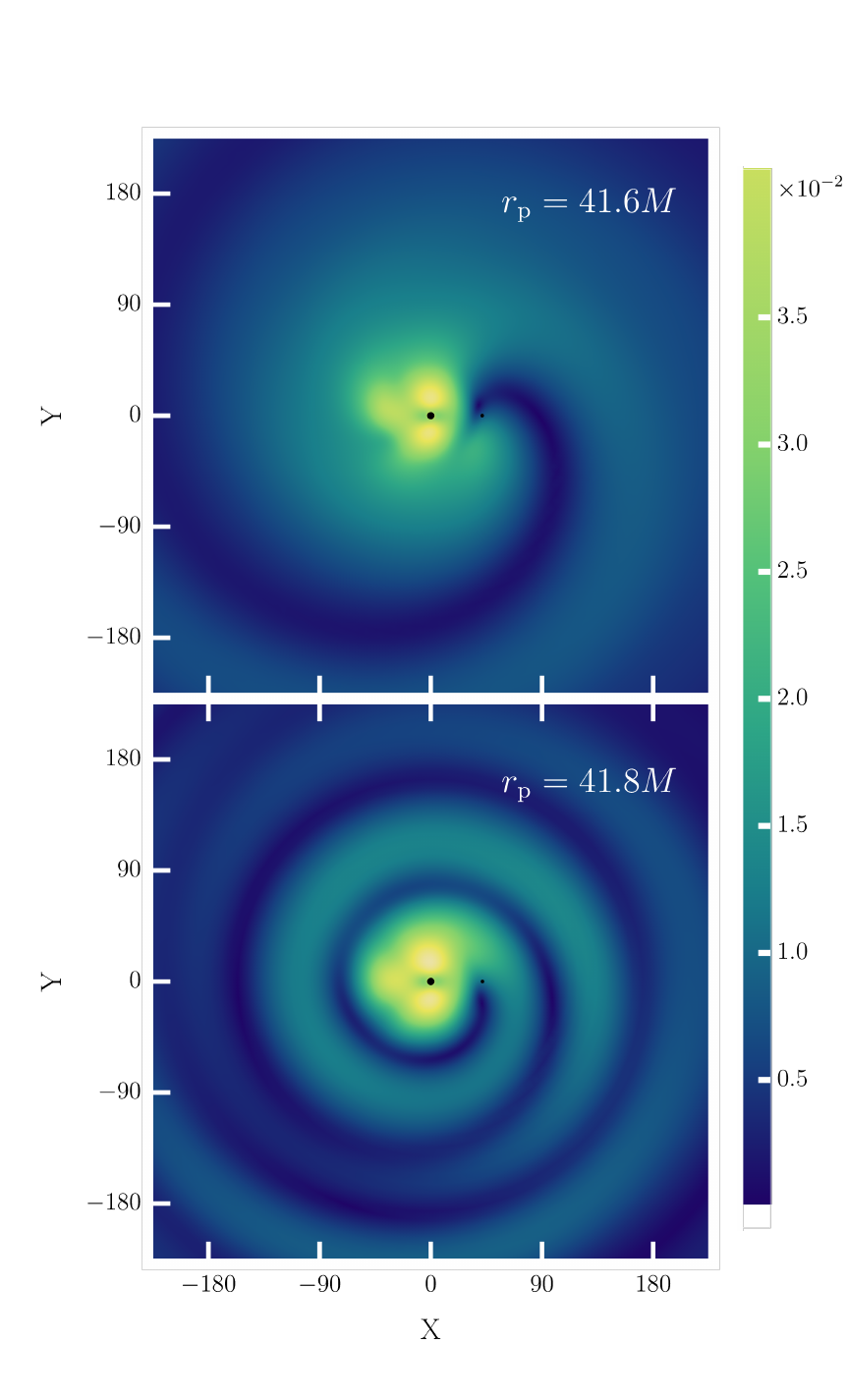}
    \caption{We show the absolute value of the perturbed scalar field $|\phi^{(1,1)}|$ for $\ell\geq 2$, taking $\alpha = 0.3$, $a = 0.88M$. In the top panel, we show an equatorial slice of the field solution at $r_{\rm p} = 41.6M$, in which the $\hat{Z}$--axis is aligned with the BH spin. In the bottom panel, we show an equatorial slice at $r_{\rm p} = 41.8M$.}
    \label{fig:Wake_Profile_SM}
\end{figure}

As discussed in the main text, the scalar fluxes to infinity exhibit sharp features~\eqref{eq:discontinuities}, as certain modes start contributing to the flux. Specifically, Eq.~\eqref{eq:discontinuities} predicts a sharp feature in the flux at $r^{*,2}_{\rm p} = 41.66M$ for $\alpha = 0.3$. In Fig.~\ref{fig:Wake_Profile_SM}, we show an equatorial slice of the field solution just before (\emph{top panel)} and after (\emph{bottom panel)} this orbital radius. Indeed, we find the morphology of the cloud to change completely ``in a short time.'' The reason being that $(\ell,m)=(2,2)$ mode of the field solution transitions from a radiative configuration---with $\sim 1/r$ decay---to a bound configuration that decays exponentially. This sharp transition in the field profile arises due to the single harmonic state configuration of the boson cloud background and will not be as prominent in environments with a more general harmonic dependence.
%
\subsection{Numerical Procedure and Validation}\label{app:numerical_procedure}
\begin{figure}
    \centering
    \includegraphics[width=\linewidth]{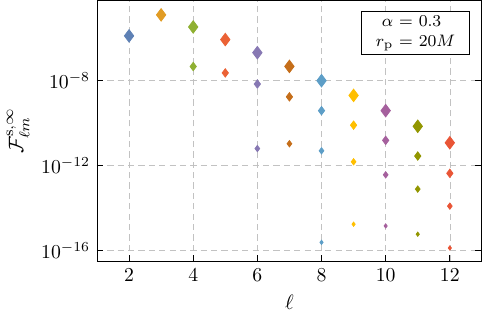}
    \caption{We show the contribution from different $\ell$-modes to the flux to infinity in Kerr for $\alpha = 0.3$ with the secondary on a prograde orbit at $r_{\rm p} = 20 M$. Due to selection rules, modes with opposite parity are zero, i.e., when $\ell$ is odd and $m$ is even or vice versa. Additionally, all $m = 0,1$ modes do not contribute to the flux to infinity. The different sized diamonds thus show the contribution from the modes that are not zero, where the higher the $m$, the higher contribution. For example, for $\ell = 8$, we show $m = 8, 6, 4, 2$. Fluxes through the horizon follow a similar trend.}
    \label{fig:flux_l_mode_convergence}
\end{figure}
\begin{figure}[!b]
    \centering
    \includegraphics[width=\linewidth]{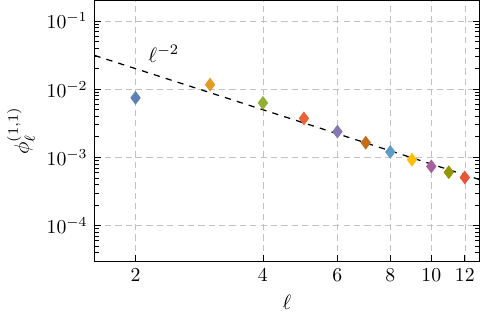}
   \caption{We show the contribution from different $\ell$-modes of the scalar field perturbation evaluated on the orbital radius of the secondary ($r_{\rm p} = 20 M$). They fall off as with the expected rate, $\ell^{-2}$, indicated by the black dashed line.}
    \label{fig:flux_l_mode_convergence_particle}
\end{figure}

To solve Eq.~\eqref{eq:EOM_sourced}, we proceed as follows.
\noindent {\bf Metric data.} To construct the metric perturbation $h^{(0,1)}_{\mu\nu}$ in Lorenz gauge, we adapt the \texttt{Mathematica} notebook developed in~\cite{Dolan:2021ijg, Dolan:2023enf, Wardell:2024yoi} to a modular \texttt{Mathematica} package suitable for exploring large parameter spaces. The package outputs spin-weighted spherical harmonic data on a tortoise coordinate grid. The data is then summed over $\ell$--modes up to $\ell_{\rm max} = 18$, constructing $m$--mode data on a two-dimensional $(r_*,\theta)$ grid. This step is essential for generating the source term in Eq.~\eqref{eq:dynamicalperubation}, as it circumvents the problem of infinite mode couplings, which would otherwise require a solution such as outlined in~\cite{Spiers:2024src}.
To check our results, the two-dimensional $m$--mode components of $h^{(0,1)}_{\mu\nu}$ are numerically reprojected onto spin-weighted spherical harmonics in the Schwarzschild limit. The resulting data is then compared against the first-order Schwarzschild Lorenz-gauge data used by the Multiscale Self-Force collaboration~\cite{Warburton:2021kwk,Wardell:2021fyy}. We find agreement at the level of machine precision across all modes and radii, with one exception:~the $(\ell,m)=(1,0)$ mode. This discrepancy is well-understood and originates from different completion choices between the two approaches. The correction term, given by Eqs.~(D3a)--(D3b) in~\cite{Miller:2020bft}, resolves this discrepancy, achieving machine precision for all $m$--modes. Importantly, the difference in the $(\ell, m) = (1,0)$ mode only affects the $m = 1$ modes of the perturbed scalar field. As the background field is in a single-state configuration with $(\ell_{\rm b}, m_{\rm b}) = (1,1)$, this mode cannot contribute to the fluxes at infinity. 
\noindent {\bf Background field.} The background scalar field $\phi^{(1,0)}$ is constructed using Leaver's method~\cite{Leaver:1985ax,Dolan:2007mj}. For a given value of the boson mass $\mu$, we construct the radial profile at the threshold frequency $\omega_{\rm c}$ of a pure $(\ell_{\rm b},m_{\rm b}) = (1,1)$ harmonic state. We then build the field profile on the same $(r_*,\theta)$ grid as the $m$--mode $h^{(0,1)}$ data. 
\noindent {\bf Derivatives.} To calculate the derivatives of the scalar field, i.e., $\nabla_\mu\nabla_\nu \phi^{(1,0)}$, we build a two-dimensional covariant derivative operator using the method of splines. As a consistency check, we contract this quantity with the background metric and confirm that $g_{\text{Kerr}}^{\mu\nu}\nabla_\mu\nabla_\nu \phi^{(1,0)} = \mu^2 \phi^{(1,0)}$ within machine precision. 
In the Schwarzchild case, the absence of a stationary threshold configuration of the cloud leads to oscillatory behaviour at the horizon, where the field oscillates as $e^{- i\omega r_*}$. This contrasts the Kerr case, where the field goes to a constant at the horizon (as $\omega \rightarrow \Omega_{\rm H}$). To ensure the robustness of our Schwarzschild results, we recompute all quantities in ingoing Eddington-Finkelstein coordinates, which do give regular derivatives at the horizon. We find that our results remain unchanged.
\noindent {\bf Source construction.} With the $m$--mode tensors $h_{(0,1)}^{\mu\nu}$ and $\nabla_\mu\nabla_\nu \phi^{(1,0)}$ validated, we contract them to form the
$m$--mode source term for Eq.~\eqref{eq:EOM_sourced}. This source term is projected onto spheroidal harmonics (or spherical harmonics in the Schwarzschild case) using the same projection routines applied earlier.
\noindent {\bf Finding solutions.} Using the radial source functions, we solve the radial Klein-Gordon equation using a variation of parameters approach. We first build the ``$\mathrm{In}$'' and ``$\mathrm{Up}$'' solution numerically as those which solve the homogenous Klein-Gordon equation with boundary conditions given by
\begin{equation}
\begin{aligned}
\lim_{r\rightarrow r_{\rm H}}R_{\mathrm{In}} &= e^{- i  (\Omega_{m_{\rm g}}+\omega_{\rm c} - m \Omega_{\rm H})r^*}\,,\\
\lim_{r\rightarrow \infty}R_{\mathrm{Up}}  &= \frac{e^{ i  k_{m_{\rm g}}  r^*} r^{\frac{i \mu^2}{ k_{m_{\rm g}}}}}{\sqrt{a^2+r^2}}\,,
\end{aligned}
\end{equation}
where $k_{m_{\rm g}} = \sqrt{(\Omega_{m_{\rm g}}+\omega_{\rm c} )^2-\mu^2}$. Taking the spheroidal projection of the source, $S^{(1,1)}_{\ell m}$,  we directly solve for the scalar field perturbations as,
\begin{equation}
    \phi^{(1,1)}_{\ell m}(r) = C^{\mathrm{Up}}_{\ell m}(r) R^{\mathrm{Up}}_{\ell m}(r) +C^{\mathrm{In}}_{\ell m}(r) R^{\mathrm{In}}_{\ell m}(r)\,,
\end{equation}
where 
\begin{equation}
\begin{aligned}
    C^{\text{Up}}_{\ell m}(r) &= \int_{r_{\rm H}}^r \frac{ R^{\mathrm{In}}_{\ell m} S^{(1,1)}_{\ell m}}{\mathcal{W}_0}\mathrm{d}r'\,, \\
    C^{\mathrm{In}}_{\ell m}(r) &= \int_{r}^\infty \frac{ R^{\mathrm{Up}}_{\ell m} S^{(1,1)}_{\ell m}}{\mathcal{W}_0}\mathrm{d}r'\,,
\end{aligned}
\end{equation}
and $\mathcal{W}_0$ is the constant Wronskian coefficient given by
\begin{equation}
   \frac{ {\mathcal{W}_0}}{\Delta}  = R^{\mathrm{In}}_{\ell m} \frac{ \mathrm{d} R^{\mathrm{Up}}_{\ell m} }{\mathrm{d}r}- R^{\mathrm{Up}}_{\ell m} \frac{\mathrm{d} R^{\mathrm{In}}_{\ell m} }{\mathrm{d}r}\,.
\end{equation}
Here, $\Delta= r^2 -2 M r + a^2$ is the standard Kerr quantity. We explicitly verify that the behavior of the solution at both boundaries aligns with the boundary conditions of the homogeneous $\mathrm{In}$ and $\mathrm{Up}$ solutions. To cross-check our code, we insert our source data in a solver from an independent implementation~\cite{Brito:2023pyl}, finding consistent results.
The solution of the Klein-Gordon equation~\eqref{eq:EOM_sourced} then gives $\phi^{(1,1)}_{\ell m}$, which we sum over $\ell$ and $m$ to reconstruct the field profile, which is shown in e.g., Fig.~\ref{fig:Wake_Profile}. The asymptotes of these extended solutions are then extracted to obtain the input of the flux formulae~\eqref{eq:scalar_flux}, generating the results shown in Fig.~\ref{fig:Fluxes_Inf_Hor}.
The amplitude of the fluxes mode-by-mode, follow the expected trend. As shown in Fig.~\ref{fig:flux_l_mode_convergence}, the ``main'' $\ell = m$ contribution decays exponentially with increasing $\ell$. Interestingly, Fig.~\ref{fig:flux_l_mode_convergence} also shows that ``sub-leading'' modes with $\ell \neq m$ can have a non-negligible contributions. For example, the $(\ell,m) = (4,2)$ mode is larger than the $(\ell,m) = (8,8)$ mode. 
Defining $\phi^{(1,1)}_{\ell} = \sum_{m=-\ell}^{\ell}\phi^{(1,1)}_{\ell m}$, we show a similar plot in Fig.~\ref{fig:flux_l_mode_convergence_particle}, now extracting the field at the position of the secondary (the point at which the field is most irregular). We find that the field perturbation is finite and continuous at the particle with the $\ell$--modes of the perturbation constituting a convergent sequence decaying as $\ell^{-2}$.
%
\end{document}